\documentclass[aps,prl,two column]{revtex4-2}
\usepackage{graphicx}
\usepackage{gensymb}
\usepackage{amsmath}
\begin{document}
\def\bb{\begin{equation}}
\def\ee{\end{equation}}

\title{Pseudospins revealed through the giant dynamical Franz-Keldysh effect in massless Dirac materials}

\author{Youngjae Kim$^{*}$}

\affiliation{
School of Physics, KIAS, Seoul 02455, Korea
}

\date{\today}

\begin{abstract}
The dynamical Franz-Keldysh effect, indicative of the transient light-matter interaction  regime between quantum and classical realms, is widely recognized as an essential signature in wide bandgap condensed matter systems such as dielectrics. In this study, we applied the time-resolved transient absorption spectroscopy to investigate ultrafast optical responses in graphene, a zero-bandgap system. We observed in the gate-tuned graphene that the massless Dirac materials notably enhance intraband light-driven transitions, significantly leading to the giant dynamical Franz-Keldysh effect compared to the massive Dirac materials, a wide bandgap system. In addition, employing the angle-resolved spectroscopy, it is found that the unique polarimetry orientation, i.e., perpendicular polarizations for the pump and the probe, further pronounces the optical spectra to exhibit the complete fishbone structure, reflecting the quantum pseudospin nature of Dirac cones. Our findings expand the establishment of emergent transient spectroscopy frameworks into not only zero-bandgap systems but also pseudospin-mediated quantum phenomena, moving beyond dielectrics.

\end{abstract}

\maketitle

In recent decades, the development of high-intensity and ultrashort laser pulses has opened new avenues for our understanding of ultrafast dynamics, enabling exploration at extreme time scales of 10 to 100 attoseconds (1 as $=$ $10^{-18}$ s)\cite{Kienberger1,Cavalieri1,Krausz1,Kruchinin,Lucchini0}. 
These advances have not only facilitated ever faster temporal resolutions within a single cycle of optical fields but also significantly achieved the study of nonlinear responses and their properties\cite{Holler1,Goulielmakis1}.

One of the central achievements is ultrafast transient absorption spectroscopy (TAS), which has emerged as a crucial technique for probing transient optical properties in various excited electronic systems. Initially originating in studies of atomic and molecular systems\cite{Goulielmakis1,Michael1,Drescher1}, TAS has been broadened to probe the characteristics of condensed matter systems: rapid dipole oscillations\cite{Mashiko1}, controls of electric carriers\cite{Schultze1,Schlaepfer1,Zurch1}, crystal orientations\cite{Sato1}, and the dynamical Franz-Keldysh effect (DFKE)\cite{Sato1,Yacoby,Jauho,Novelli1,Otobe1,Du1,Lucchini1,Lucchini2,Volkov}.

The DFKE becomes completely characterized in the transient regime between quantum and classical realms\cite{Novelli1,Otobe1}, particularly when non-interacting wide bandgap systems, such as dielectrics, are exposed to intense and alternating external electric fields.
This effect is mainly marked by distinct features\cite{Lucchini2,Picon,Sato1,Otobe1,Dong,Srivastava}: real-time V-shaped phase oscillations, known as the fishbone structure induced by intraband transitions, and the redshift in their differential TAS. 
The redshift originates from field-induced modifications of the band structures, as interpreted in the Franz-Keldysh effect (FKE)\cite{Franz,Keldysh,Tharmalingam,Nahory,Otobe1}, leading to an increase in absorption spectra below the bandgap edge in the adiabatic limit.

Recent experimental and theoretical studies have expanded the understanding of the DFKE. For instance, an experimental studies on diamond have demonstrated the transient absorptions induced by intense pump pulses, also clearly explained by the first-principles calculations\cite{Lucchini1,Lucchini2}. Further theoretical works have explored many-body interacting systems beyond the DFKE. In multi-band charge transfer insulators within the time-dependent density functional theory plus U, the DFKE is observed to be screened due to the strong screenings in many-body correlations\cite{Dejean1}. Moreover, in single-band Mott insulating phases computed by the time-dependent exact diagonalizations, nontrivial aspects of the DFKE are induced by the transitions of double occupancy without screenings\cite{Kim1}. 
However, the exploration of transient dynamics via TAS has been still predominantly confined to wide bandgap systems. Given these advancements, it is now timely to propose expanding the establishments of TAS to another avenue, unveiling emergent transient dynamics beyond the dielectrics.

On the other side, graphene, considered as an essential quantum material, exhibits a range of intriguing properties, from nonlinear optical effects\cite{Yoshikawa,Sato2,Jiang,Kim2} to topological\cite{KWKim,McIver,Oka,Kim3}. A key aspect of graphene is the presence of massless Dirac fermions, categorizing it as a zero-bandgap system. Considering this point, we focus on the gate-tuned graphene. The gate-tuning can modulate the chemical potential to control its optical responses where the interband transitions of Dirac cone become forbidden due to empty states above the chemical potential\cite{Li,Horng} whereas the intraband transitions are still allowed. Although a gate-tuned graphene was previously suggested to exhibit the DFKE\cite{Zhou}, deep demonstrations under the ultrafast TAS have been rarely studied, primarily because the FKE was rooted in dielectrics\cite{Kruchinin}.

In this study, we demonstrate the time-resolved TAS to investigate emergent optical responses and transient dynamics in gate-tuned graphene. We find that the signal of the DFKE evolves distinctively with a given chemical potential. Moreover, angle-resolved polarimetry orientations for the pump and probe pulses reveal that the absorption responses vary with the angle. Eventually, perpendicular polarizations predominantly yields much more explicit DFKE signals, characterized by the fishbone structures and the redshifts, reflecting the pseudospin nature of graphene.

In addition, comparing the massless and the massive Dirac materials under identical laser pulse conditions, we observe that the DFKE in the massless Dirac systems is significantly pronounced, by the enhanced intraband light-driven transitions\cite{Picon} due to its massless electronic structures. Our findings propose that gate-tuned graphene would be an ideal platform for the emergent researches for the DFKE, extending the framework of transient spectroscopy into zero-bandgap and pseudospin-mediated quantum phenomena.

\begin{figure}
\includegraphics[width=0.48\textwidth]{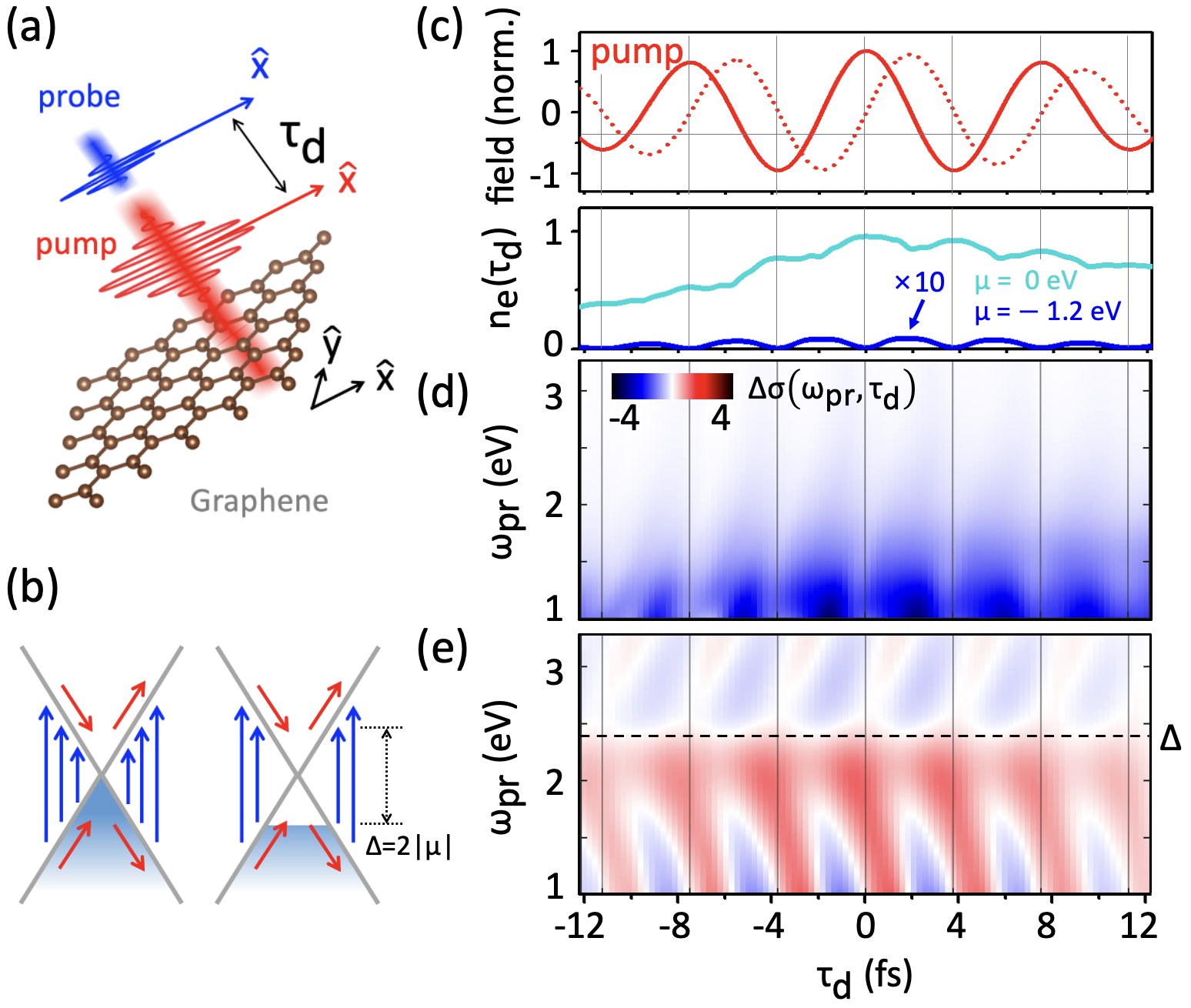}
%\vspace*{7.5cm}
%\special{psfile=fig1.png}
\caption{
The ultrafast TAS on gate-tuned graphene.
(a), Schematics of the pump and probe spectroscopy for the TAS. The pump and the probe pulses are aligned with same direction, i.e., the x-direction.
(b), Dirac cones with the chemical potential alternation, $\mu$, which introduces $\Delta = 2|\mu|$. The red and blue arrows stand for optical transitions of the pump and the probe, respectively. 
(c), Normalized time-profile of the pump pulse. The peak electric field is $E_{0} = 0.04 V/\AA$. The solid and the dotted lines represent the vector potential and the electric field, respectively. The bottom inset shows the time-dependent excited electrons $n_{e}$ in arbitrary unit, in cases of $\mu = $0 eV and $-$1.2 eV.
(d),(e), Calculated ultrafast TAS $\Delta\boldsymbol{\sigma}(\omega_{pr},\tau_{d})$ for $\mu = $0 eV (d) and $-$1.2 eV (e).
}
\label{FIG1}
\end{figure}

We employ the time-dependent equation of motion to elucidate the microscopic quantum mechanical light-matter interactions in Dirac materials\cite{Sato2}, expressed as,
\begin{equation}
{\partial}/{{\partial}\tau}\rho_{\textbf{k}}(\tau)={-i}[H_{\textbf{k}(\tau)},\rho_{\textbf{k}}(\tau)]
\end{equation}

Here, the density matrix $\rho_{\textbf{k}}(\tau)$, defined within the Houston states describing the instantaneous electronic states at momentum $\textbf{k}(\tau)$. The Hamiltonian $H_{\textbf{k}(\tau)}$ is expanded from momenta $\textbf{K}$ and $-\textbf{K}$ and can be described as $H_{\textbf{k}(\tau)}=v_{F}\xi\bar{\sigma}_{x}k_{x}(\tau)+v_{F}\bar{\sigma}_{y}k_{y}(\tau)$, where $\bar{\sigma}_{x(y)}$ denotes the Pauli matrix. The Fermi velocity $v_F$ in graphene is given in the energy unit of $-$13.88 eV, and $\xi$ is the valley parameter for $\textbf{K}$ ($\xi=1$) and $-\textbf{K}$ ($\xi=-1$), and both $\xi$ are included in our calculation. The time-dependent momentum $\textbf{k}(\tau)=\textbf{k}-\textbf{A}(\tau)/c$ is derived using the Peierls substitution, with the speed of light $c$ and the vector potential $\textbf{A}(\tau) \rightarrow \textbf{A}_{pu}(\tau)+\textbf{A}_{pr}(\tau-\tau_{d})$: the pump and the probe pulses, respectively, as depicted in the Fig.\ref{FIG1}(a). The pump pulse is $\textbf{A}_{pu}(\tau)=A_{0}\cos(\omega_{0}\tau)\cos^{2}(\pi\tau/{2\tau_0})\hat{\textbf{x}}$ at $|\pi\tau/{2\tau_0}|<\pi/2$ and $\textbf{A}_{pu}(\tau)=0$ elsewhere. We adopt the ultrashort (a few cycles) NIR pump pulse with 27 fs of $\tau_0$ and 0.55 eV of $\omega_{0}$, thus $\sim$ 7 optical cycles are included in a single pulse packet. The probe pulse is ${A}_{pr}(\tau) = \lambda\cos(\Omega_{pr}\tau)\cos^{2}(\pi\tau/{2\tau_{pr}})$ at $|\pi\tau/{2\tau_{pr}}|<\pi/2$ and ${A}_{pr}(\tau)=0$ elsewhere, with ultrashort parameters $\Omega_{pr}$ of 2.4 eV and $\tau_{pr}$ of 1.7 fs. The $\lambda = 0.001|v_{F}|$ for the linear response limit. The probe is polarized along the x-direction unless stated otherwise, i.e., $\textbf{A}_{pr}(\tau) = {A}_{pr}(\tau)\hat{\textbf{x}}$ (also see the Fig.\ref{FIG1}(a)).

As depicted in Fig.\ref{FIG1}(b), the initial distribution of electrons is thermalized by the chemical potential $\mu$, follows by $\rho_{\textbf{k}}(\tau \rightarrow -\infty )_{n,m} = \delta_{n,m}(e^{(\varepsilon_{\textbf{k},n}-\mu)/k_{B}T}+1)^{-1}$, with temperature $T =$ 80 K, the Boltzmann constant $k_{B}$, and the initial eigen energies $\varepsilon_{\textbf{k},n}$.
Therefore, the dynamics in Dirac cones vary with $\mu$. 
At $\mu =$ 0 in the charge neutral states, both inter- and intraband transitions contribute to the dynamics. When $\mu$ attains a finite value in the gate-tuned states, however, interband transitions below $\Delta = 2|\mu|$ become forbidden, while intraband transitions continue to occur.
Fig.\ref{FIG1}(c) shows the temporal profile of the pump laser pulse applied to graphene, with the peak electric field strength $E_{0} =$ 0.04 $V/\AA$ of $\textbf{E}_{pu}(\tau)=-(1/c)\partial \textbf{A}_{pu}(\tau)/\partial \tau$.
The bottom inset of Fig.\ref{FIG1}(c) displays the number of excited electrons as a function of $\mu$. The number of excited electrons is $n_{e}(\tau) = 1/(2\pi)^2\int d\textbf{k}^2[\rho_{\textbf{k}}(\tau)_{n,n}-\rho_{\textbf{k}}(\tau \rightarrow -\infty )_{n,n}]$ with $n$ for the upper Dirac cones. In case of charge neutral state, the electrons are significantly excited due to the zero-bandgap structures of Dirac cone. In the gate-tuned states with $\mu = -$1.2 eV, however, direct optical excitations are not permissible so that $n_{e}$ can be ignored, emphasizing the remaining intraband transitions. Eventually, in the gate-tuned graphene, the pump pulse mostly contributes to the intraband transitions, and similarly, the interband transitions are mainly contributed from the probe pulse.

To investigate time-resolved TAS using transient optical conductivities, first we obtain the optical conductivities of initial states as,
\begin{equation}
\boldsymbol{{\sigma}}_{0}(\omega_{pr}) = {\Re}[\textbf{J}_{pr}(\omega_{pr})/\textbf{E}_{pr}(\omega_{pr})]
\end{equation}
Here, ${\Re}[...]$ extracts the real part in the $[...]$. The current density $\textbf{J}_{pr}(\tau)$ is obtained from $\textbf{J}_{pr}(\tau)=1/(2\pi)^2\int d\textbf{k}^2 tr[\textbf{j}_\textbf{k}(\tau)\rho_{\textbf{k}}(\tau)]$, with the current operator $\textbf{j}_\textbf{k}(\tau)=-\partial H_{\textbf{k}(\tau)}/\partial \textbf{A}(\tau)$ and $\textbf{k}(\tau)=\textbf{k}-\textbf{A}(\tau)/c$ for $\textbf{A}(\tau)=\textbf{A}_{pr}(\tau)$. These values are transformed into frequency space as $\textbf{J}_{pr}(\omega) = \int d\tau e^{-\eta\tau+i\omega\tau} \textbf{J}_{pr}(\tau)$ and $\textbf{E}_{pr}(\omega) = \int d\tau e^{-\eta\tau+i\omega\tau} \textbf{E}_{pr}(\tau)$, with $\eta = 0.3$ eV as a damping parameter.

Subsequently, we calculate transient conductivities by $\textbf{J}_{tr}(\tau,\tau_d)=\textbf{J}_{{pu}+{pr}}(\tau,\tau_d)-\textbf{J}_{{pu}}(\tau)$. The $\textbf{J}_{{pu}+{pr}}(\tau,\tau_d)$ should be the current densities when the Dirac hamiltonian $H_{\textbf{k}(\tau)}$ is evolved by the vector potential $\textbf{k}(\tau)=\textbf{k}-\textbf{A}(\tau)/c$ for $\textbf{A}(\tau)=\textbf{A}_{pu}(\tau)+\textbf{A}_{pr}(\tau-\tau_{d})$ and similarly, $\textbf{J}_{{pu}}(\tau)$ under $H_{\textbf{k}(\tau)}$ for $\textbf{A}(\tau)=\textbf{A}_{pu}(\tau)$. 
Finally, the difference spectra of TAS, which we shall call TAS hereafter, is practically obtained as\cite{Sato1},
\begin{equation}
\Delta \boldsymbol{\sigma}(\omega_{pr},\tau_d) =
{\Re}[\textbf{J}_{tr}(\omega_{pr},\tau_d)/\textbf{E}_{pr}(\omega_{pr},\tau_d)]-
\boldsymbol{\sigma}_{0}(\omega_{pr})
\end{equation}

Fig.\ref{FIG1}(d) and (e) depict TAS results to compare charge neutral states ($\mu =$ 0) with gate-tuned states ($\mu =-$1.2 eV). In the charge neutral states (Fig.\ref{FIG1}(d)), the TAS shows negative signs continuously across the entire spectra, a consequence of Pauli blocking by excited electrons and holes, as indicated in the inset of Fig.\ref{FIG1}(c). In contrast, Fig.\ref{FIG1}(e) shows that the TAS of gate-tuned graphene exhibits positive signs at $\omega_{pr} <$ 2.4 eV where the $\Delta = 2|\mu| =$ 2.4 eV for $\mu=-$1.2 eV. Above 2.4 eV, the TAS shows dominant negative signals. These results correspond to the redshifts, one of the signatures of the DFKE\cite{Lucchini2,Otobe1}. However, the fishbone structure, another signature of the DFKE, are not completely observed at this stage, despite the explicit presence of time-dependent phase oscillations with frequency of 2$\omega_0$.  
We note that time-resolved spectroscopy on sub-laser-cycle electron dynamics necessitates $\omega_{pr} \gg \omega_0$, limiting our discussion to low frequency range at $\omega_{pr} <$ 1 eV.

\begin{figure}
\includegraphics[width=0.48\textwidth]{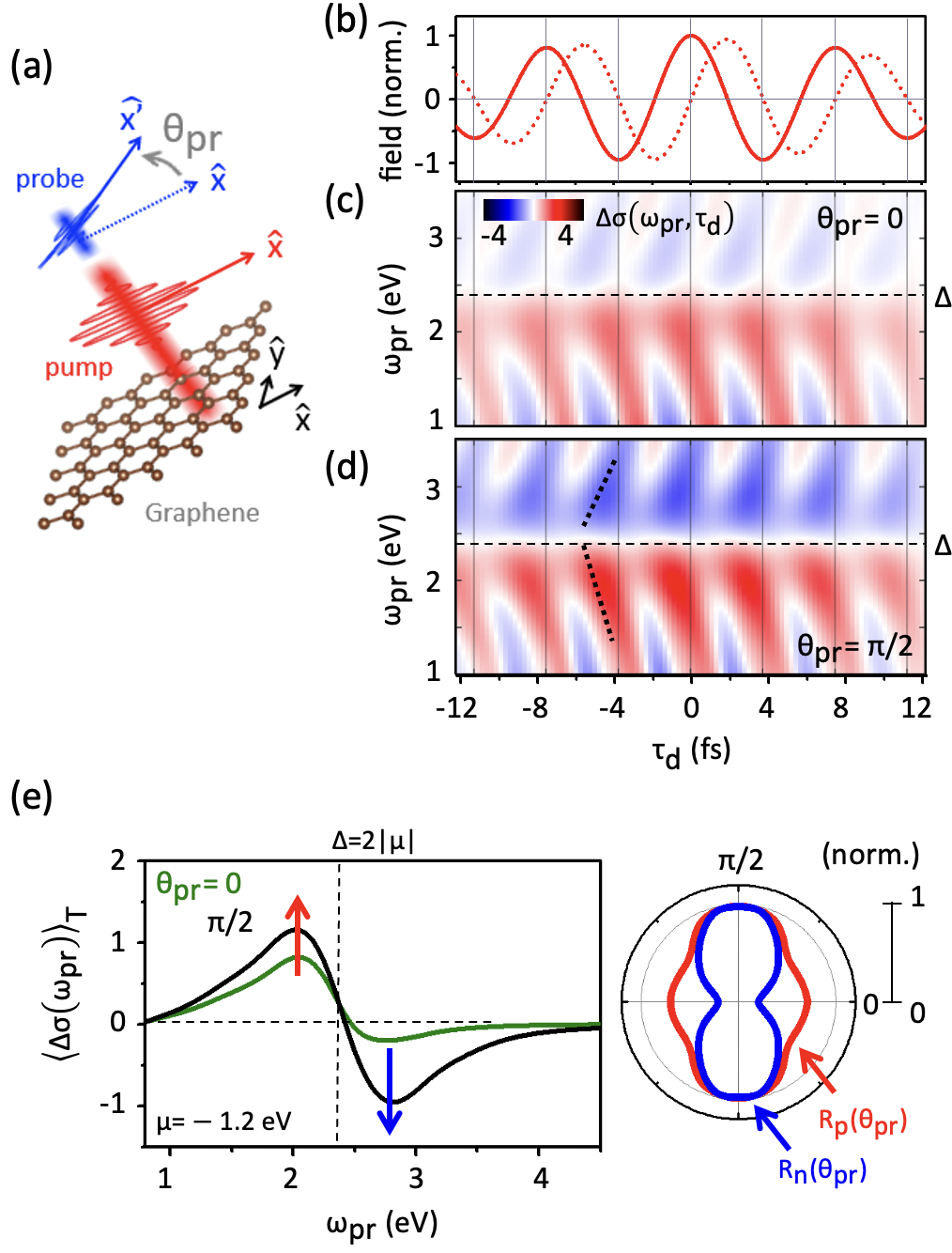}
%\vspace*{7.8cm}
%\special{psfile=fig2.png}
\caption{
Calculated TAS on the gate-tuned graphene with angle-resolved polarimetry orientations between the pump and the probe pulses.
(a), Schematics of the angle-resolved geometry. The polarization of the probe pulse, $\textbf{x}'$, is rotated by $\theta_{pr}$ with respect to the x-direction.
(b), Normalized time-profile of the pump pulse with $E_{0} = 0.04 V\AA$
(c),(d), Calculated TAS at the $\theta_{pr}=$ 0 (c) and the $\theta_{pr}=\pi/2$ (d).
The dotted line in the (d) is drawn as the guide for eyes for the fishbone structure.
Note, spectra of (c,d) correspond to the $\mu=-$1.2 eV. 
(e),(f), Calculated cycle-averaged TAS, $\langle \Delta \boldsymbol{\sigma}(\omega_{pr})\rangle_{T} = 1/T_{0}\int_{-T_{0}/2}^{T_{0}/2} d\tau_d \Delta \boldsymbol{\sigma}(\omega_{pr},\tau_d)$ where time-period of pump pulse, $T_{0}=2\pi/\omega_{0}$. 
The cycle-averaged TAS in cases of $\mu=-$1.2 eV (e) as a function of $\theta_{pr}$. In the polar plot inset shows the characteristic polarimetries in the spectra defined as normalized $R(\theta_{pr})$, that is, $R_{p}(\theta_{pr})$ (red line) and $R_{n}(\theta_{pr})$ (blue line). The $R_{p}$ and $R_{n}$ denote for highest positive and lowest negative values of $\langle \Delta \boldsymbol{\sigma}(\omega_{pr})\rangle_{T}$, respectively, i.e., $R_{p}(\theta_{pr})=max[\langle \Delta \boldsymbol{\sigma}(\omega_{pr})\rangle_{T}]$ and $R_{n}(\theta_{pr})=|min[\langle \Delta \boldsymbol{\sigma}(\omega_{pr})\rangle_{T}]|$ at given ${\theta_{pr}}$.
}
\label{FIG2}
\end{figure}

In Fig.\ref{FIG2}, based on angle-resolved polarimetry orientations between the pump and the probe, we further demonstrate how the TAS $\Delta \boldsymbol{\sigma}(\omega_{pr},\tau_d)$ changes to figure out an incomplete signal when the polarization of the probe pulse is rotated to align along $\hat{\textbf{x}}'$, i.e., $\hat{\textbf{x}}' = \cos\theta_{pr}\hat{\textbf{x}} + \sin\theta_{pr}\hat{\textbf{y}}$. 
Consequently, the probe pulse is set as $\textbf{A}_{pr}(\tau) = {A}_{pr}(\tau)\hat{\textbf{x}}'$, while the pump pulse $\textbf{A}_{pu}(\tau)$ remains as $\textbf{A}_{pu}(\tau) \parallel \hat{\textbf{x}}$, as shown in Fig.\ref{FIG2}(a). We now investigate the $\Delta \boldsymbol{\sigma}(\omega_{pr},\tau_d)$ along $\hat{\textbf{x}}'$. As illustrated in Fig.\ref{FIG2}(b-d), it is notable that the observed spectra depend on the angle $\theta_{pr}$, even under the same pump pulse characteristics. Interestingly, at $\theta_{pr} = \pi/2$ (Fig.\ref{FIG2}(d)), the spectra distinctly reveal the missing feature of the DFKE in Fig.\ref{FIG1}, that is, the fishbone structure, as indicated by the dotted line. This suggests that an unusual orientation, $\theta_{pr} = \pi/2$, between the pump and probe is crucial in the case of massless Dirac systems to measure the complete signal of the DFKE.

To identify the rotation-dependent behaviors in the spectra, we present the cycle-averaged spectra, defined over a single period of the pump pulse in Fig.\ref{FIG2}(e). The rise and the decay behaviors of the spectra below and above $\Delta$ $(=2|\mu|)$, respectively, are clearly observed for various $\theta_{pr}$ values. These behaviors coincide with the DFKE in wide bandgap systems and can be interpreted as the effect of the FKE\cite{Otobe1,Seraphin}. Notably, the decay behaviors of the spectra at $\omega_{pr} > \Delta$ exhibit a more dramatic response with respect to $\theta_{pr}$ compared to the rise behaviors at $\omega_{pr} < \Delta$, which is more clearly visible in the insets of polar plots in Fig.\ref{FIG2}(e),(f).
In the characteristic polarimetry of the inset, it is also evident that $\theta_{pr}=\pi/2$ is a critical point where the lowest negative value of $\langle \Delta \boldsymbol{\sigma}(\omega_{pr})\rangle_{T}$ is comparable to its highest positive value. This indicates that the entire spectra become symmetric forms around $\Delta$, corresponding to pronounced intraband light-driven transitions completely detectable by the TAS.

Moreover, the DFKE in gate-tuned graphene is found to clearly exist even under strong bath couplings since the decoherence does not affect the DFKE, where the intraband light-driven transitions play an essential role (see Fig. S1 in the SM\cite{Sato2,Sato3}).

\begin{figure}
\includegraphics[width=0.48\textwidth]{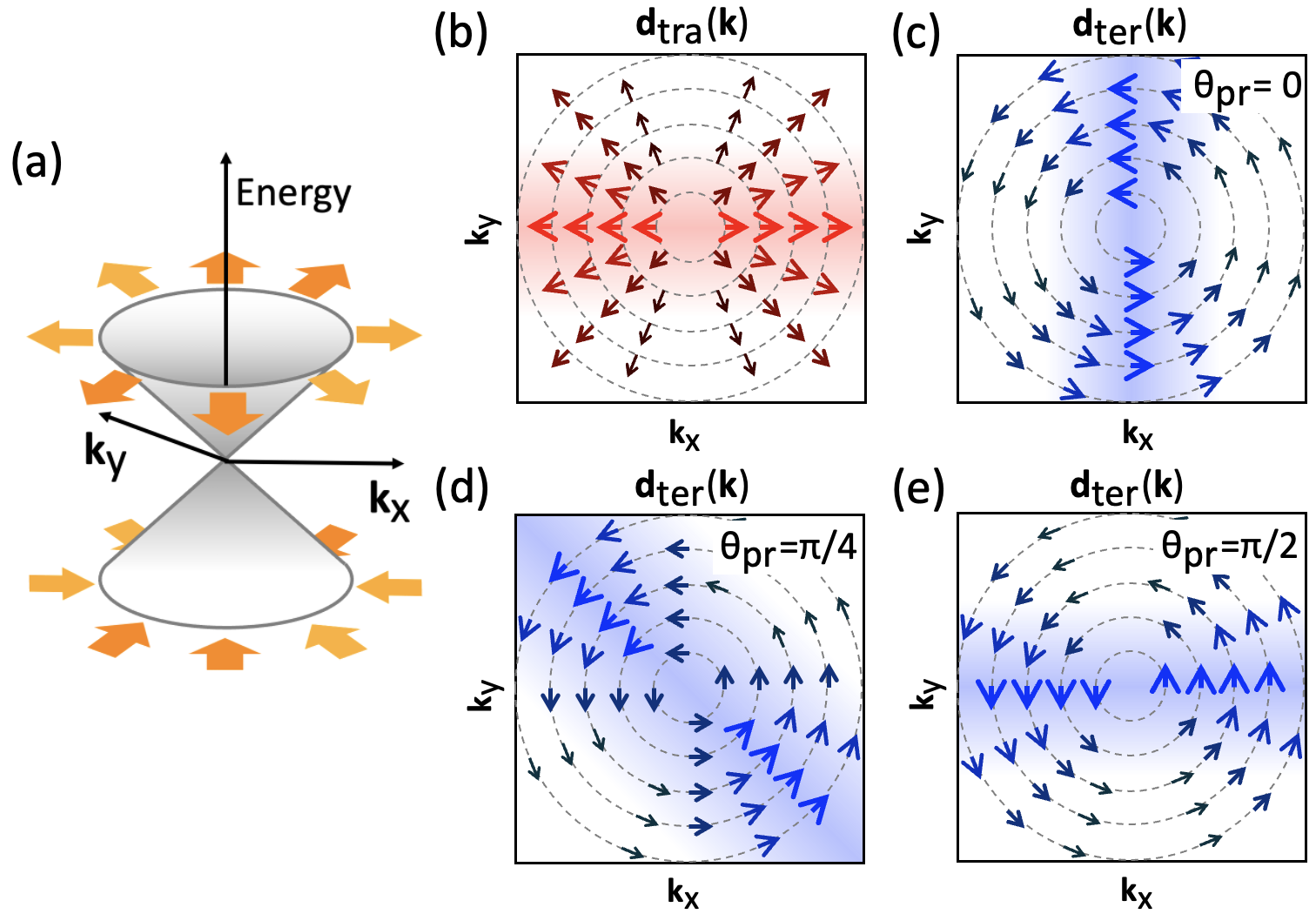}
%\vspace*{7.8cm}
%\special{psfile=fig4.png}
\caption{
Quantum pseudospin textures and related transition momentum matrix.
(a), In-plane arranged pseudospin textures (yellow arrows) in a Dirac cone.
(b-e), Normalized momentum matrix of a Dirac cone displayed in the momentum space. 
Note, the dashed concentric circles represent the isoenergy lines for Dirac cone centered at the Dirac point. 
The size of surrounding arrows and their colors indicate the strength of momentum matrix under given laser field direction. The red and the blue colors correspond to the intraband and the interband elements, respectively.
Each arrow indicates the direction of momentum matrix.
(b), Intraband momentum matrix $\textbf{d}_{tra}(\textbf{k})$ under the x-polarized laser field, i.e., the pump pulse. (c-e), Interband momentum matrix $\textbf{d}_{ter}(\textbf{k})$ under the laser field, i.e., the probe pulse, under various $\theta_{pr}$: 0 (c), $\pi$/4 (d), and $\pi$/2 (e).
}
\label{FIG3}
\end{figure}

To gain a deeper understanding of the physics underlying massless Dirac materials, we examine the unique electronic structures of the Dirac cone, as depicted in Fig.\ref{FIG3}. In graphene, due to the identical sub-lattice degree of freedom, the Dirac cone can be expressed by the pseudospin notation\cite{Echtermeyer} using the Pauli matrix $\bar{\sigma}$ (see Fig.\ref{FIG3}(a)). Consequently, the momentum matrix can be formulated as $\textbf{A}\cdot\bar{\boldsymbol{\sigma}}$. The intraband momentum matrix $\textbf{d}_{tra}(\textbf{k})$ and the interband momentum matrix $\textbf{d}_{ter}(\textbf{k})$ are represented as $\textbf{d}_{tra}(\textbf{k})=\langle\pm\textbf{k}|\textbf{A}\cdot\bar{\boldsymbol{\sigma}}|\pm\textbf{k}\rangle$ and $\textbf{d}_{ter}(\textbf{k})=\langle\pm\textbf{k}|\textbf{A}\cdot\bar{\boldsymbol{\sigma}}|\mp\textbf{k}\rangle$, respectively, where $|n\textbf{k}\rangle$ corresponds to the eigenstates of the Dirac cone.

When a x-polarized laser is applied, the $\textbf{d}_{tra}$ predominantly arises from the $k_{x}$ sectors of the Dirac cone, as illustrated in Fig.\ref{FIG3}(b), which corresponds to the pump pulse in our study. In contrast, $\textbf{d}_{ter}$ behaves differently due to its pseudospin properties. As shown in Fig.\ref{FIG3}(c), when a x-polarized laser is applied, $\textbf{d}_{ter}$ aligns perpendicularly along the $k_{y}$ sectors, which mainly corresponds to the probe pulse in our study. In this case, the pump and probe pulses do not fully overlap in momentum space. This explains why incomplete spectra of the DFKE are observed in parallel polarization between the pump and probe pulses, as shown in Fig.\ref{FIG1} and Fig.\ref{FIG2}.  However, as depicted in Fig.\ref{FIG3}(d) and (e), as the orientation of the probe pulse rotates to increase $\theta_{pr}$, the two momentum matrices begin to overlap. Notably, the $\theta_{pr} = \pi/2$ results in the full overlap and yields complete spectra of the DFKE with fishbone structures, as evidenced in Fig.\ref{FIG2}, reflecting the critical role of quantum pseudospin nature in massless Dirac systems.
The schematic in Fig.\ref{FIG3}(a) represents a single $\xi$, and that the momentum matrices in Fig.\ref{FIG3}(b-e) are identical for both $\xi$ values.
The results of pseudospins in this study are different with the x-ray attosecond TAS in graphene\cite{Cistaro} where the x-ray probe photons excite electrons from core states to upper Dirac cones states, so that pseudospin selection rules are not encountered. In other words, optical transitions between the lower and upper Dirac cones are important for capturing the pseudospin selection rules.

\begin{figure}
\includegraphics[width=0.48\textwidth]{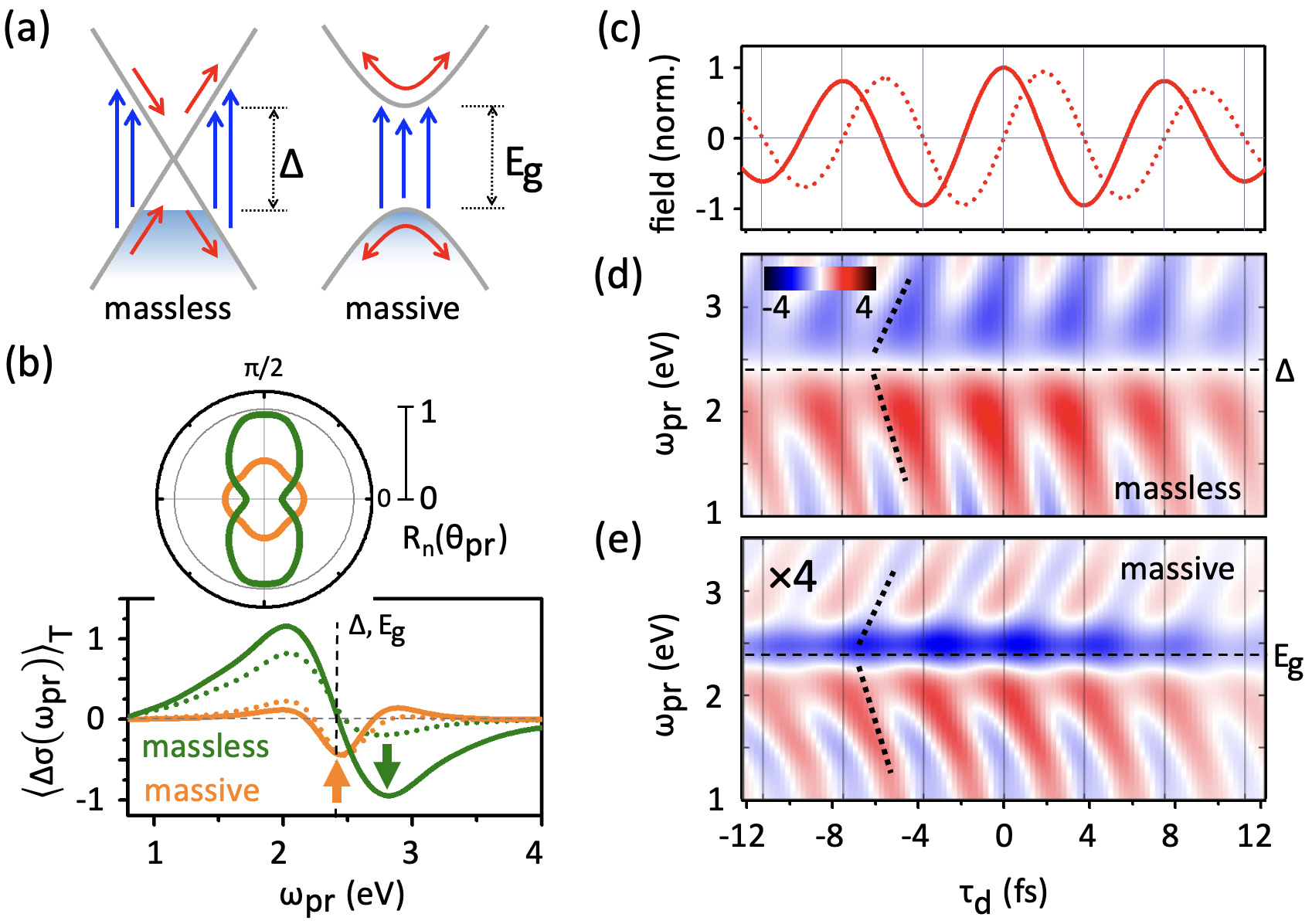}
%\vspace*{7.8cm}
%\special{psfile=fig4.png}
\caption{
The giant DFKE measured in the massless Dirac materials and a comparison between the massless and the massive Dirac materials.
(a), Schematics of dynamics in massless Dirac materials with $\mu = -1.2$ eV, $\Delta=2|\mu|$ and the massive Dirac materials with band gap $E_{g}$ of 2.4 eV.
(b), The cycle-averaged TAS of massive and massless Dirac materials. The solid and dotted lines represent the results of $\theta_{pr} = \pi/2$ and $\theta_{pr} = 0$, respectively.
The arrow symbols indicate the minimum value of each TAS.
The polar plot inset shows $\theta_{pr}$-dependent minimum value of spectra, $|R_{n}(\theta_{pr})|$, ($R_{n}(\theta_{pr})=min[\langle \Delta \boldsymbol{\sigma}(\omega_{pr})\rangle_{T}]$) (see the arrows), for the massless and the massive Dirac materials.
(c), Time-profile of the pump pulse. $E_{0} = 0.04 V\AA$.
(d),(e), Calculated TAS for the massless Dirac materials with $\theta_{pr} = \pi/2$ (d) and the massive Dirac materials with $\theta_{pr} = 0$. Note, the (d) is zoomed-in image about four times.
}
\label{FIG4}
\end{figure}

To discuss the DFKE of graphene moving beyond dielectrics, we present a comparative analysis of TAS between graphene and wide-bandgap systems, examining both massless and massive Dirac materials in Fig.\ref{FIG4}. To incorporate massive Dirac materials, we introduce a mass parameter $E_{g}$ into the Dirac Hamiltonian used in Eq.(1), by $H_{\textbf{k}(\tau)} \rightarrow H_{\textbf{k}(\tau)} + E_{g}/2\bar{\sigma}_{z}$. This allows us to compare dynamics between massless Dirac materials with $\mu=-$1.2 eV ($\Delta =$ 2.4 eV) and massive Dirac materials with $E_{g}=$2.4 eV under identical laser conditions, as shown in Fig.\ref{FIG4}(a).

In Fig.\ref{FIG4}(b), we observe distinct behaviors in the cycle-averaged TAS for both systems.
We emphasize two remarkable aspects in this figure. 
First, in massless Dirac materials, the spectra at $\omega_{pr} < \Delta$ exhibit pronounced redshifts, overwhelming those in massive Dirac systems across all $\theta_{pr}$ values. This originates from the high velocity of massless Dirac materials, enhancing intraband light-induced transitions\cite{Picon} compared to massive systems. 
Second, in massless Dirac materials, a dramatic decrease in negative spectra at $\omega_{pr} > \Delta$ is observed as the $\theta_{pr}$ increases, as indicated by the green arrow in Fig.\ref{FIG4}(b). In contrast, the massive Dirac systems show no significant change under varying $\theta_{pr}$, as shown by the yellow arrow.
The polar plot inset of Fig.\ref{FIG4}(b) clearly illustrates this phenomenon. In the inset, the signal of the massless Dirac material is varying with $\theta_{pr}$, and pronounced at $\theta_{pr}=\pi/2$. This demonstrates the pseudospin nature of the Dirac cone, which exhibits in-plane pseudospins in the entire momentum space, as depicted in Fig.\ref{FIG3}(a). In contrast, massive Dirac systems show constant responses regardless of $\theta_{pr}$, due to their pseudospins aligning along the z-direction at the bandgap edge, lacking in-plane directional properties.

Additionally, the Fig.\ref{FIG4}(c,e) display the TAS results for both systems, clearly illustrating the DFKE, characterized by fishbone structures and redshifts. Although the TAS from the massive systems (Fig.\ref{FIG4}(e)) conventionally represent the DFKE, we note that the TAS from massless systems (Fig.\ref{FIG4}(d)) shows much pronounced DFKE features, suggesting graphene as an ideal material for studying the DFKE.

\begin{figure}
\includegraphics[width=0.48\textwidth]{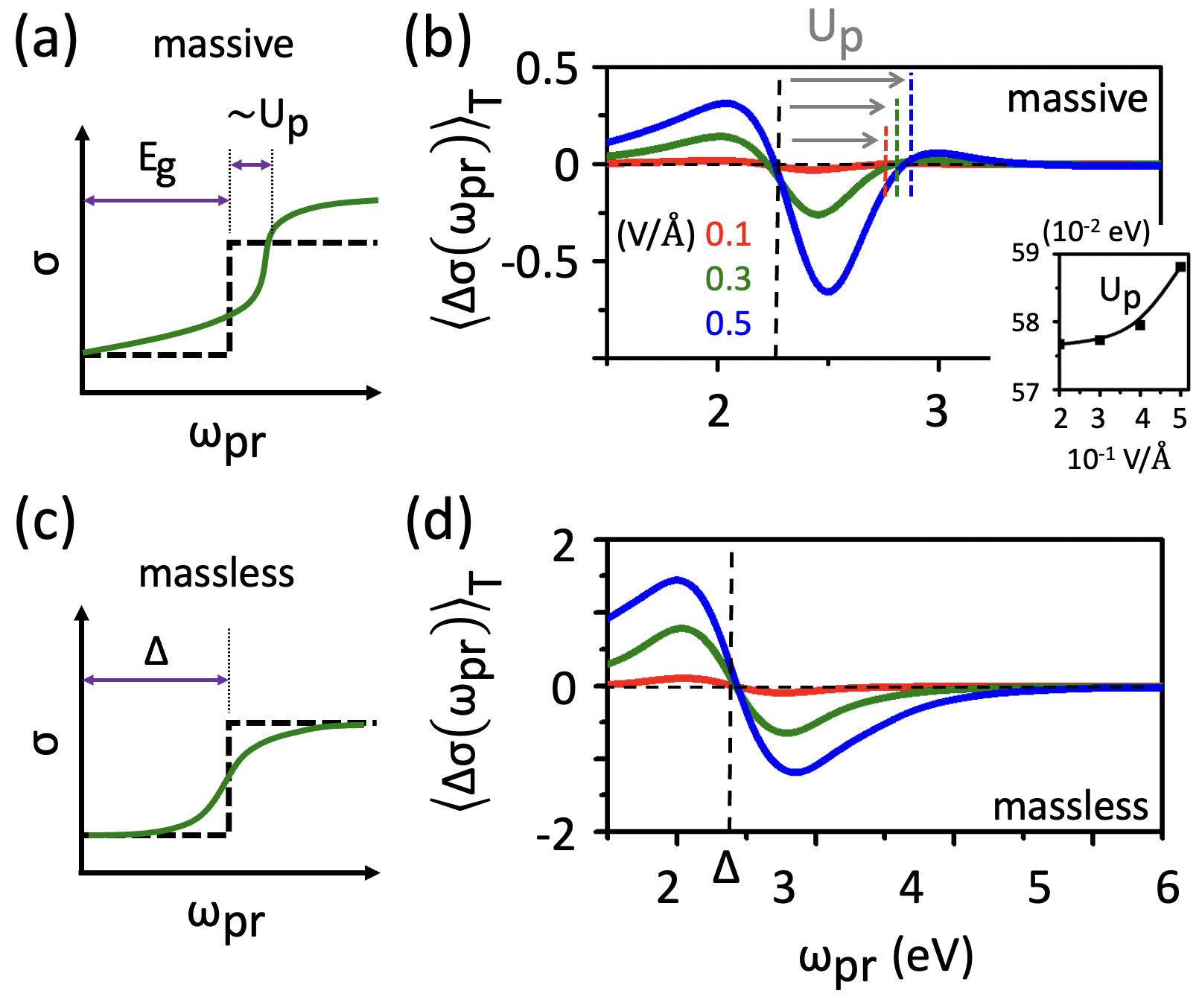}
%\vspace*{7.5cm}
%\special{psfile=fig5.png}
\caption{
The ponderomotive shift of the massive (a,b) and the massless Dirac systems (c,d).
(a), Schematics of ponderomotive shift $U_{p}$ in massive Dirac systems of unperturbed absorption (dashed black line) and dressed absorption (solid green line). 
(b), Cycle-averaged TAS with different pump intensities. The inset displays estimated $U_{p}$ as a function of pump field strength.
(c), Schematics of ponderomotive shift in massless Dirac systems. The $U_{p}$ is not observed.
(d), Cycle-averaged TAS with different pump intensities.
The gap parameters are equally set at $\Delta=E_{g}=$ 2.4 eV,
}
\label{FIG5}
\end{figure}

In our last discussion of the ponderomotive shift between massive and massless systems, as shown in Fig.\ref{FIG5}, we observed critical responses in the massless Dirac systems.

Under intense laser pump driving, the electronic structures become dynamically dressed, resulting in an increased effective bandgap, which reflects the ponderomotive shift as shown in Fig.\ref{FIG5}(a). This shift has already been well demonstrated in gapped systems \cite{Lucchini2,Volkov,Srivastava2}. The calculated cycle-averaged TAS in massive systems, under various pump intensities, also exhibits the ponderomotive shift, as displayed in Fig.\ref{FIG5}(b). The inset shows a quadratic dependence on the field strength, consistent with the definition of the ponderomotive energy, $U_{p} = \frac{A_{0}^{2}}{2\mu c^{2}}$, where $\mu$ is the reduced mass.

However, for massless systems, one might expect the ponderomotive shift to diverge as $\mu \rightarrow 0$. This is not observed due to the failure of the quadratic form of $U_{p}$. For massless Dirac systems, the ponderomotive energy should be redefined as $U_{p} = \frac{1}{T_{0}}\int_{0}^{T_{0}} d\tau (\varepsilon_{\textbf{k}-\textbf{A}(\tau)/c,n}-\varepsilon_{\textbf{k}-\textbf{A}(0)/c,n})$\cite{Kruchinin}, where the massless Dirac band is defined by $\varepsilon_{\textbf{k},n}=v_Fk$. Consequently, $U_{p} = 0$ due to the linear term. Thus, as shown in Fig.\ref{FIG5}(c) and (d), the cycle-averaged TAS does not exhibit the ponderomotive shift, which leads to a refined description of the DFKE decoupled from the ponderomotive shift in massless Dirac materials, a distinct behavior from that observed in dielectrics.

In summary, our study on time-resolved TAS in graphene reveals distinct optical responses and the evolution of the DFKE with gate-tuning. Angle-resolved polarization orientations for the pump and probe pulses significantly affect the TAS signals, especially when aligned perpendicularly, identifying the pseudospin nature of Dirac cones. A comparative analysis between massless and massive Dirac materials under identical laser conditions shows the pronounced DFKE in massless systems, attributable to high intraband transitions of the Dirac cones. These findings suggest that gate-tuned graphene is an ideal platform for DFKE studies, extending the TAS framework into zero-bandgap systems and offering insights into the real-time quantum pseudospin dynamics, thereby advancing the potential realm of ultrafast spectroscopy throughout condensed matter physics.

\section*{Acknowledgement}
The authors thank the computational support from the Center for Advanced Computation (CAC) at Korea Institute for Advanced Study (KIAS).
In this study, Y.K. supported by a KIAS Individual Grant (PG088601) at Korea Institute for Advanced Study (KIAS).
\\ \\
${}^{\ast}$ Corresponding author: ykim.email@gmail.com

\end{document}